# "I'm happy even though it's not real": GenAI Photo Editing as a Remembering Experience

Yufeng Wu
School of Computer Science
University of Technology Sydney
Sydney, NSW, Australia
yufeng.wu-2@student.uts.edu.au

Qing Li
School of Computer Science
University of Technology Sydney
Sydney, NSW, Australia
qing.li-6@student.uts.edu.au

Elise van den Hoven
Materialising Memories & Visualisation Institute
University of Technology Sydney
Sydney, NSW, Australia
elise.vandenhoven@uts.edu.au

A. Baki Kocaballi
School of Computer Science
University of Technology Sydney
Sydney, Australia
baki.kocaballi@uts.edu.au

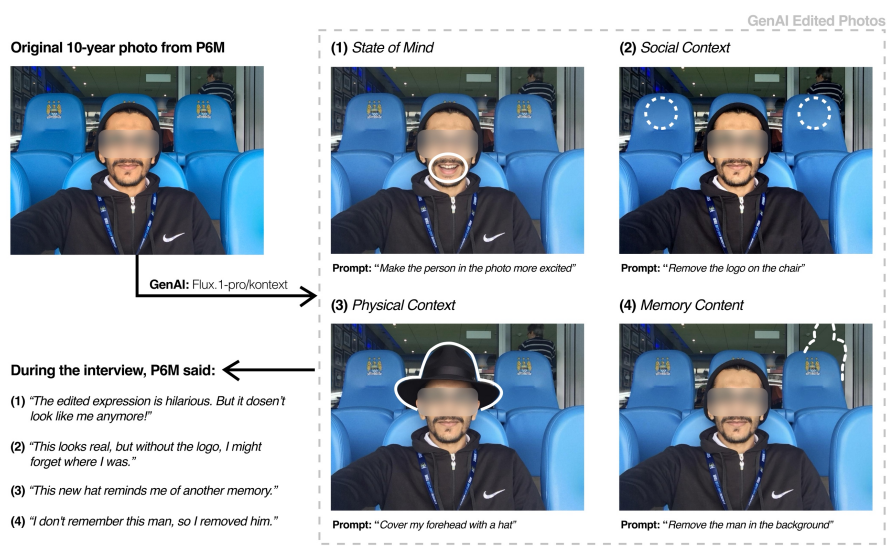

Figure 1: An example of the GenAI editing process from our study. The figure shows participant P6M's journey from an original photo (top left) to four edited photos (right), each guided by a different Remembering Experience (RX) dimension. These dimensions provide participants with scenarios to guide their edits based on State of Mind (internal feelings), Social Context (sharing with others), Physical Context (physical display), and Memory Content (matching memory details). The participant's interview reflections show how this editing became part of the remembering experience itself.

## Abstract

Generative Artificial Intelligence (GenAI) is increasingly integrated into photo applications on personal devices, making editing photographs easier than ever while potentially influencing the memories they represent. This study explores how and why people use GenAI to edit personal photos and how this shapes their remembering experience. We conducted a two-phase qualitative study with 12 participants: a photo editing session using a GenAI tool guided by the Remembering Experience (RX) dimensions, followed by semi-structured interviews where participants reflected on the editing process and results. Findings show that participants prioritised felt memory over factual accuracy. For different photo elements, environments were modified easily, however, editing was deemed unacceptable if it touched upon a person's identity. Editing processes brought positive and negative impacts, and itself also became a remembering experience. We further discuss potential benefits and risks of GenAI editing for remembering purposes and propose design implications for responsible GenAI.

**ACM Reference Format:**
Yufeng Wu, Qing Li, Elise van den Hoven, and A. Baki Kocaballi. 2026. "I'm happy even though it's not real": GenAI Photo Editing as a Remembering Experience. In *Proceedings of the 2026 CHI Conference on Human Factors in Computing Systems (CHI '26), April 13–17, 2026, Barcelona, Spain.* ACM, New York, NY, USA, 18 pages. https://doi.org/10.1145/3772318.3790927

## 1 Introduction

Imagine if one day a friend disappeared from every photo you had together. Much like how Hermione Granger in *Harry Potter* used magic to erase herself from her parents' photographs, would you still remember your friend if all visual traces were gone? Photographs have long served as a "prosthetic memory," a medium that anchors personal and collective recall [13, 42]. However, the proliferation of consumer-facing Generative Artificial Intelligence (GenAI) tools, such as Google Photos' Magic Eraser [86], Apple Intelligence's edit functions [87], and Adobe's generative fill [88], has altered photographs' role. Photo editing, once the domain of professionals, is now a routine part of everyday media practices [59]. These technologies now go far beyond simple filters or red-eye removal, allowing users to seamlessly add, remove, or modify any elements within their personal photos, shifting photographs from static records to malleable media.

This shift has raised concern in HCI and cognitive science about misinformation and eroding trust [72], and GenAI-edited media can implant false memories and distort recollections [59]. Yet the emphasis on deception overlooks emerging, self-authored practices in everyday remembering. People increasingly use GenAI for "new modes of expression, conversation, creativity, and ways of overcoming forgetting" [25], not to fabricate, but to align subjective, emotional, and sensory remembering with the photographic record. However, we still lack empirical evidence explaining how and why people use GenAI to edit personal photographs, as well as how they perceive and respond to its potential impacts on remembering. Such insights can inform the design of responsible GenAI that support meaning-making while also sustaining the foundations of trust.

We adopt the Remembering Experience (RX) concept, which frames remembering as a holistic, situated activity rather than a matter of factual recall. Prior work [29] defines the RX as "the set of effects that is initiated during the situated recall of a personal past episode," and identify four contextual and cognitive dimensions that shape it: State of Mind, Social Context, Physical Context, and Memory Content. In HCI, RX has been used to study and design technologies for remembering that foreground affective, embodied, and social contexts of recall [26, 28, 56]. A binary true or false memory framework is not useful to understand how people use GenAI to modify personal photos in everyday life and how such practices affect their personal RX [69]. In this study, we operationalised each RX dimension into an open-ended question to guide participants' photo-editing tasks, encouraging them to reflect within different remembering contexts.

We conducted a two-phase explorative qualitative study with 12 participants that involved (1) a think-aloud photo editing session where participants used a state-of-the-art GenAI to modify personal photos from three different time points in the past, guided by four RX dimensions, and (2) photo-elicitation interviews. This study was guided by the following research questions:

RQ1: Why and how do people use GenAI to edit personal photos?

RQ2: What photo elements are seen as editable vs. inviolable, and why?

RQ3: How does photo age, and the memory it represents, shape editing practices and perceptions?

Our analysis shows that participants prioritised the subjective and affective dimensions of their recollections over documentary accuracy. In line with reconstructive views of memory, we introduce *felt memory* to describe participants' focus on edits that felt congruent with their subjective experience of past events [74, 81]. We found that this pursuit was driven by *intentions* such as amplifying emotion and clarifying narrative. To enact these intentions, participants employed a range of *Edit Types* (modification, addition, removal, and extension) directed at key *Edit Targets* (environment, human, text, and meta). A *hierarchy of editability* emerged: faces as identity anchors; conditional acceptance for appearance and body edits based on plausibility; and environments as malleable, "safe" contexts. Furthermore, photo age shaped these editing strategies, descriptive patterns suggest that older photos attracted slightly more active reconstruction. Editing itself became a form of active remembering that could both enrich the experience, yet risking information loss and persuasive shifts in recollection.

Based on these findings, we contribute an integrated account of how GenAI photo editing is becoming part of everyday remembering experiences, and why this matter for both personal meaning-making and the credibility of personal archives. First, we offer felt memory as a lens for understanding GenAI-mediated photographic remembering beyond documentary accuracy, and for articulating how edits can be oriented toward experiential resonance. Second, we propose a hierarchy of editability as a vocabulary for describing the boundaries people set when editing personal photos, distinguishing identity-critical elements from context-oriented changes. Third, we translate these insights into design directions for responsible GenAI photo editing that support expressive remembering while remaining attentive to risks around trust and distortion in revisiting and sharing personal media.

## 2 Related Work

Research has shown how photographs can mediate remembering in everyday life, and more recently how GenAI is transforming the role of photographs. Yet, while much attention has gone to misinformation and false memory, we still know little about the lived experience of engaging with GenAI-edited personal photographs. Our work addresses this gap.

## 2.1 Remembering through Photographs

Much of the early cognitive psychology literature framed remembering in terms of encoding and retrieval processes [1, 10], although other studies already emphasised its reconstructive nature [2]. More recently, HCI has increasingly framed remembering as a situated experience encompassing affective, embodied and social dimensions [26, 28, 29, 56]. While some research has explored how tangible interactions [17, 24, 48] and AR [67] can support this experience, personal photographs consistently remain the most ubiquitous and socially powerful medium for everyday remembering [56].

Consequently, a significant body of HCI research focuses on how people engage with photos to reconstruct memory. Foundational work by Kirk et al. introduced the notion of "PhotoWork," which refers to the activities people perform with digital photos after capturing but before sharing, such as reviewing, editing, and organising [31]. Importantly, PhotoWork is not simply the administrative handling of images but a form of memory work because selecting, arranging, and revisiting photos actively shapes how events are understood and remembered over time. Later research expanded this perspective through the concept of "PhotoUse," demonstrating that these activities are not merely functional but purposive, driven by specific autobiographical goals such as reminiscing, social bonding, and identity formation [6].

Design research further demonstrates how interacting with stored photos can deepen reflection and memorialisation. Systems such as *Photobox* [54] and *Story Shell* [78] create conditions for slow, serendipitous, or ritualised engagement, revealing how different modes of encountering photos influence how the past is revisited. Social and communicative perspectives similarly show that photos function as cues for storytelling [4] and identity construction, enabling people to negotiate how their experiences are represented to others [9, 63, 68]. Historically, this body of work has conceptualised PhotoWork primarily in terms of curation, referring to the process of choosing which moments to keep, display, or archive to represent the past [30]. Within this framing, editing has typically been treated as a corrective or aesthetic step.

## 2.2 The Transformation of Photographs in the GenAI Era

This shift toward construction is driven by how GenAI fundamentally alters the nature of the photographic medium. Photographs have long functioned as a form of prosthetic memory, valued for their indexical relationship to reality and their role as an anchor of identity and emotion [13, 42]. Traditional photo manipulation was possible before GenAI, but it remained largely constrained to professionals or skilled users employing tools such as Photoshop [88]. Everyday editing typically focused on corrective or cosmetic adjustments, while substantial semantic changes were rare. Studies of beautification in social media similarly show that most consumer editing involved enhancing appearance (e.g., smoothing skin or adjusting facial features) rather than reconstructing scenes, reinforcing the widespread cultural assumption that personal photographs were relatively stable records of the past [57].

Recent advances in diffusion-based [44] and transformer-based models [15] have converted photographs from static records into malleable media. By allowing users to seamlessly generate or modify content through natural-language prompts [43], these tools fundamentally lower the barrier for non-experts to alter the core semantics of a photo. In this shift, photographs move from being primarily objects of curation to materials that can be reshaped, expanded, or reinterpreted, marking a transition from selection to reconstruction within everyday photo practices.

## 2.3 The Role of GenAI in Remembering Experiences

GenAI editing now permeates everyday photo use through three common forms: (a) semantic edits that add, remove, or alter people, objects, and settings, shifting an image's meaning (e.g., changing ethnicity or time of day) [79], (b) minimal "enhancements" that promise better quality without overt semantic change [60], and (c) image-to-video animation that turns still photos into short, plausible clips [35, 36]. These affordances are increasingly surfaced in mainstream apps and even default camera workflows (e.g., Google's Best Take [85]), and are applied in recent work (e.g., Sora [89], Dream Machine [90], Kling [91]), further blurring boundaries between capture and synthesis [16, 59].

Despite their rapid integration into everyday tools, very little research has examined how these affordances shape user practices. Practitioner analyses of GenAI-based photo restoration suggest that even minimal "enhancements" can homogenise facial features and subtly rewrite visual records, raising questions about how such tools shape what is preserved or remembered in an image [11]. For semantic edits and image-to-video animation, Pataranutaporn et al. [59] experimentally demonstrated that GenAI-edited images, especially GenAI-generated videos derived from them, significantly increase false memory. Classic studies have long shown how suggestion, photographs, and other visual manipulations can induce false recollections [22, 39, 40], and recent work extends this agenda to AI-synthesised and deepfaked media [49, 51].

Although remembering experience research has long emphasised technologies as external memory cues [28], and photographs as powerful triggers for recall [46], little is known about how GenAI-edited personal photos shape everyday remembering practices. We lack an empirical account of self-authored GenAI photo editing as a remembering practice, including people's *intentions*, *editing practices*, and *experienced outcomes*. This gap motivates our study of how everyday editors use GenAI tools to align photographs with how moments are remembered, and where they draw lines of acceptability.

More broadly, GenAI transforms personal photographs from relatively stable records of past events into malleable materials that can be revised to fit how moments are felt, narrated, and socially communicated [57]. As editing becomes easy and visually persuasive, this shift may reshape behaviour and perception by changing what people treat as acceptable transformation, how confidently they rely on photos when revisiting events, and how edited photos are interpreted when shared beyond their original context, where editing intent and history may not travel with the photo. In this sense, GenAI photo editing can become part of active remembering, with the potential to enrich meaning-making while also introducing conditions for subtle distortion, source confusion, and eroding trust in personal archives and everyday media [21, 76].

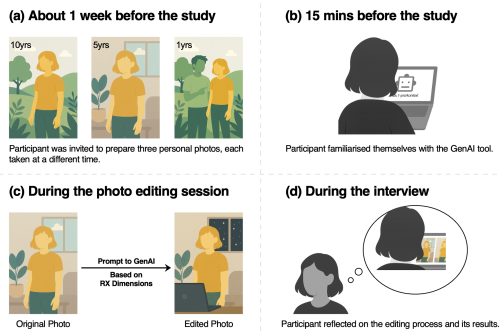

Figure 2: An overview of this two-phase study procedure. The process involved (a) participant preparation of personal photos, (b) familiarisation with the GenAI tool, (c) a think-aloud editing session guided by RX dimensions, and (d) a semi-structured interview for reflection on the editing photos and process.

These implications motivate the need to articulate responsible GenAI photo editing beyond a narrow focus on malicious deception. Drawing on responsible and trustworthy AI perspectives, we use responsible GenAI photo editing to refer to designs and practices that preserve user agency and oversight, promote transparency and accountability, respect privacy and data governance, and manage risks through robustness and ongoing assessment [12, 92]. In the context of remembering experiences, these framing foregrounds keeping transformations interpretable across time and audiences, for example by making provenance and intent legible and by supporting reflective, context-sensitive editing decisions, so that GenAI can support expressive remembering without inadvertently undermining the credibility that personal photos often carry.

## 3 Method

We conducted a qualitatively led, two-phase study to elicit rich, contextualised accounts of how RX shaped participants' GenAI photo-editing practices, and how the resulting edits in turn reshaped their RX. Phase 1 captured in-the-moment reflections during guided editing sessions; Phase 2 used photo-elicitation interviews to support retrospective sense-making. We report descriptive statistics (counts, proportions, simple summaries of ratings) to map what was edited and how, and to triangulate the qualitative analysis. These numeric summaries are within-sample descriptors only; no inferential statistics were performed given the small, purposive sample.

### 3.1 Participants

We recruited 12 participants from Australia via social media and community forums for in-person sessions (Table 1). The sample was stratified to ensure diversity across different demographics. Participants were balanced by gender (six female, six male) and by self-reported familiarity with GenAI (six none-to-low, six moderate-to-high). Familiarity was defined by how often participants used GenAI in daily life: none (never), low (tried once or twice), moderate (occasional use), and high (regular use). The age range was 24–54 (M=39.9), evenly distributed across three cohorts: 20–35 (n=4), 36–50 (n=4), and 51–65 (n=4). Inclusion criteria required participants to provide personal photos from one, five, and ten years prior and be comfortable with reflective discussion. The inclusion of photos from different time spans allowed us to explore whether editing practices varied with the age of the memory represented. Participants are referenced as P# plus gender (e.g., P1F = Participant 1, Female).

Table 1: Participant Demographics and GenAI Familiarity

| ID | Age | Gender | Familiarity |
|----|-----|--------|-------------|
| 1  | 26  | Female | Moderate    |
| 2  | 26  | Male   | High        |
| 3  | 31  | Female | Low         |
| 4  | 24  | Male   | High        |
| 5  | 54  | Female | Low         |
| 6  | 40  | Male   | High        |
| 7  | 52  | Female | None        |
| 8  | 38  | Female | None        |
| 9  | 36  | Male   | Low         |
| 10 | 51  | Male   | High        |
| 11 | 56  | Male   | Low         |
| 12 | 45  | Female | High        |

### 3.2 Using the RX Dimensions to Guide Editing

To guide the editing tasks, we operationalised the four RX dimensions [29] into open-ended questions. This approach ensured that our data collection was clearly structured and theoretically grounded while remaining sensitive to the lived RX. Each question framed a distinct remembering context, encouraging participants to reflect within different aspects of remembering while ensuring the tasks were both theoretically grounded and sensitive to the lived RX. We identified the core focus (e.g., emotional state, social and audience context, physical setting, and match between remembered details and the record) and rephrased it into a single, plain-language question. The wording was iteratively refined through team discussions and further refined by two pilot sessions based on participant feedback to ensure that the key ideas of each RX dimension were understandable to participants. These prompts were intended as reflective scaffolds in a qualitatively led study, not as a psychometric scale, and we therefore did not perform a separate validation. For each of their three photos, participants could make up to four edits, each linked to one RX dimension:

**State of mind:** "Think about how you felt at the moment this photo was taken. Would you like to make any changes to better reflect how you felt?" This dimension captures how the individual's feelings and emotions at the time of the event shape the RX.

**Social context:** "Imagine you are holding this photo and sharing its memory with others. Would you like to make any changes

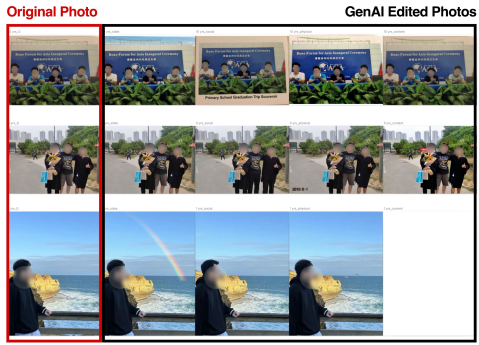

Figure 3: Digital canvas displaying original photos (left) and GenAI edited photos (right) during the interview. The bank rectangle indicates a skipped RX Task.

before sharing this memory?" This dimension probes how social presentation and audience shape RX.

**Physical context:** "Envision seeing this photo and reminiscing in a different setting—say, framed on your wall or in a photo album. Would you like to make any changes?" This dimension explores how physical settings influence the RX.

**Memory content:** *"Would you like to make any changes, so the photo better matches your memory of the event?"* This dimension examines the remembered details and emotions and their impact on the RX.

### 3.3 GenAI Tool

Before selecting *Flux.1-pro/kontext* [93], three of the authors conducted small pilot editing sessions with stock images (n=4) to compare available GenAI photo-editing tools, including Adobe Photoshop (with generative edit function) [88], Google AI Studio [94], and Open AI based image generation models [84, 95]. For each tool we tested a common set of operations (object removal, background extension, relighting, and subtle face edits) via natural-language prompts. We accessed Flux.1-pro/kontext through the web-based fai.ai platform [96], and judged it to be the most suitable for our purposes because (1) it consistently followed conversational prompts for precise local edits, (2) produced visually coherent outputs, and (3) met our privacy requirements by not retaining or retraining on user images. This process was intended only to ensure the tool was adequate for our tasks, not as a systematic benchmark.

### 3.4 Data Collection

Each participant completed the study in a single session of approximately 90 minutes, conducted in person at University of Technology Sydney. The study was approved by the university's ethics committee (application number: ETH25-10819, CCS Internal Project Number: 2025-3). All participants signed a consent form, agreeing to screen and audio recording, and authorising the use of their photos in this research. The overall procedure refers to Figure 2.

**Pre-session preparation (15 minutes):** Before the study, participants were contacted one week in advance and asked to select three personally meaningful photos from ~1, 5, and 10 years ago, each with at least one identifiable person and a moment of memory significance. They chose the photos freely from their own libraries and brought the digital files to the session. At the start, the researcher explained the study, obtained consent, clarified data handling, introduced the four RX dimensions, after which participants used a stock image to familiarise themselves with the GenAI.

**Phase 1: photo editing session (30–40 minutes):** The first phase was a think-aloud photo editing session. Using a university laptop, participants edited each photo across up to four RX Tasks, with each task guided by the corresponding open-ended question. Participants first chose which of the three photos to edit and in what order, based on their personal preference. For each selected photo, the order of the four RX Tasks was determined by a predefined randomised schedule generated in a spreadsheet to prevent order effects. Participants were presented with all four tasks but could skip any task if they preferred not to make edits. Sessions were screen- and audio-recorded. After completing their edits, participants rated each edited photo on two 5-point Likert scales: one for "Realness" and one for "Emotional Connection."

**Realness:** Compared to the original photo, how realistic or believable the edited photo appeared (1 = not at all realistic, 3=neutral, same as the original, 5 = even more realistic than the original).

**Emotional connection:** Compared to the original photo, how strongly the edited photo evoked an emotional connection to the memory (1 = not at all, 3=neutral, same as the original, 5 = even stronger than the original).

**Phase 2: semi-structured interview (40-60 minutes):** After a short break (~10 minutes), the researcher presented each participant's original and edited photos on a digital canvas (Figure 3) as elicitation material. The semi-structured interview guide was developed by the team to align with our research questions and the RX dimensions operationalised in Phase 1 and was iteratively refined through internal piloting to improve clarity, ordering, and pacing. The guide was organised into three parts: (1) a structured walk-through of each original photo and its edited versions, focusing on what was changed, why, and how the edit related to the participant's remembering experience; (2) comparative reflection across versions, including perceived authenticity and emotional impact and where participants drew boundaries of acceptability; and (3) broader questions about perceived benefits, concerns, and sharing considerations of GenAI photo editing in everyday remembering, with open-ended questions used throughout to elicit concrete examples and minimise demand characteristics.

### 3.5 Data Analysis

To analyse the data, we first defined key terms (see Table 2). Each RX Task corresponded to one of the four RX dimensions and was associated with one Prompt and one Edited Photo. Although participants could revise their prompt multiple times within the same RX Task, our observations showed that these revisions always expressed the same underlying intention. Participants only made small wording adjustments (e.g., "make the sky brighter" → "increase brightness of the sky"), and none of them introduced two semantically different prompts within a single task. Because these revisions did not add new meaning or change the resulting edit, we coded only the final prompt for each RX Task.

Table 2: Key terms and outcome categories

| Term | Definition |
| --- | --- |
| RX Task | One of the four RX dimensions (State of Mind, Social Context, Physical Context, Memory Content), each guiding participants to enter a prompt. |
| Skipped RX Task | The participant chose not to enter any prompt for a given RX Task. |
| Prompt | The text instruction entered by a participant into the GenAI for a given RX Task. |
| Edited Photo | The resulting image generated from the participant's final Prompt for a given RX Task. |
| Edit Request | A single change specified within a prompt (e.g., "add sunshine" or "remove a stranger"); one prompt may contain multiple edits. Includes both successful and unsuccessful edits. |
| Intention | The motivation participants articulated in the post-editing interviews for making a specific Edit Request; each Edit Request is linked to one Intention |
| Unsuccessful Edit | The GenAI producing irrelevant or mismatched changes, or no visible change. |

Within each Prompt, every Edit Request was treated as the core unit of analysis. For example, the prompt "add sunshine and remove a stranger" was coded as two Edit Requests. Each Edit Request was linked to one Intention articulated in the interviews. Intentions were thus counted at the level of Edit Requests rather than Prompts. For each Edit Request, we coded its type and target and linked it to the intention (from the interview), completion status (successful or not), and participants' ratings of *Realness* and *Emotional Connection* (from Likert scale). Unsuccessful Edits were also coded.

**Data cleaning:** All photos were removed from the platform after being securely downloaded to our local storage system, where they were de-identified to protect participant privacy.

**Content analysis:** All participant-entered prompts were logged, and each edit request was coded by *Edit Type* (e.g., modification, removal) and *Edit Target (*e.g., human, environment) to develop a taxonomy of editing behaviours [32].

**Reflexive thematic analysis:** We analysed participants' intentions and reflections using reflexive thematic analysis [5]. The first and second authors independently coded all think-aloud and interview transcripts. Coding was inductive and iterative, with codes progressively organised into candidate themes through discussion and refinement. The full author team regularly reviewed these themes to surface alternative interpretations, resolve overlaps, and test coherence across the dataset. In line with a reflexive stance, we acknowledge that our disciplinary backgrounds shaped interpretation. The first author brought expertise in memory and GenAI, the second in GenAI, the third in photography and memory, and the fourth in photography and GenAI. These perspectives influenced what we attended to in the data and how we evaluated its salience. Candidate themes were reviewed against the complete set of coded extracts and transcripts to ensure internal coherence and distinctiveness. Illustrative quotes were selected to convey the breadth of perspectives, including cases that complicated or challenged dominant patterns (e.g., participants who used edits to create a fantastical or surreal image). Such cases were retained to sharpen theme boundaries and highlight diversity in experience.

**Descriptive statistical analysis:** Frequencies of edit types, targets, intentions, and completion status, together with Realness and Emotional Connection ratings, were analysed descriptively to identify patterns and triangulate the qualitative findings.

## 4 Findings

We first describe participants' intentions and editing practices, their perceptions of the edited photos, and the relationship between edit types and targets (Section 4.1). We then explore how the age of the photo influenced these practices (Section 4.2). Finally, we integrate thematic analysis with earlier findings to explain participants' broader motivations, the impact of editing on their remembering experience, and their attitudes toward GenAI photo editing (Section 4.3).

### 4.1 From Intention to Edit

During the photo-editing sessions, participants generated 105 Prompts, resulting in 105 Edited Photos. These Prompts contained 133 Edit Requests (including 16 Unsuccessful Edits), each linked to a corresponding Intention. In addition, out of 144 total RX Tasks, 39 were skipped, yielding no Edit Requests or Intentions. To contextualise these skipped tasks, we examined their distribution across RX dimensions. Out of 39 skipped tasks, most skips occurred in Memory Content (n=15) and State of Mind (n=12). In these cases, participants noted they either lacked a vivid memory or felt that the photo already aligned closely with their memory. In contrast, Physical Context (n=7) and Social Context (n=5) were skipped less frequently.

*4.1.1 Intention Analysis.* To understand *why* participants want to edit their photos, we performed reflexive thematic analysis [5] on interview transcripts. This analysis revealed eight distinct themes. Table 3 presents these themes with their definitions and frequencies.

This distribution highlights the diversity of participants' intentions. While the most common intentions centred on amplifying emotions, clarifying narratives, and adjusting appearances, other themes such as distraction removal, memory cue addition, and creative expression also emerged.

*4.1.2 Edit Requests Analysis.* To understand *how* and *what* participants chose to edit, we conducted a content analysis of the 133 Edit Requests. Two coding schemes were developed directly from the data: one for Edit Type (the action performed) and one for Edit Target (the object of the action). The full taxonomies and frequencies are detailed in Tables 4 and 5

**Edit Type:** Our analysis shows that participants predominantly chose to change existing content (modification, n=62). Creative

Table 3: Analysis results of intentions for photo editing, showing theme definitions and frequency.

| Theme | Definition | Frequency |
| --- | --- | --- |
| Emotional Amplification | Edits intended to enhance or intensify the mood or feeling. | 29 (21.8%) |
| Narrative & Contextual Clarity | Edits aimed at making the story or context of the memory easier for others to understand. | 25 (18.8%) |
| Appearance Adjustments | Edits focused on altering the appearance of subjects in the photo, such as clothing, facial features, hairstyle to align with personal standards. | 22 (16.6%) |
| Distractions Refinement | Edits intended to remove or minimise irrelevant elements (e.g., secondary individuals, unwanted objects). | 17 (12.8%) |
| Memory Cue Addition | Edits aimed at adding elements that were not existing the photograph, serving as triggers for memory. | 14 (10.5%) |
| Memory Alignment | Edits aimed to make to the photograph more accurately correspond with the participant's felt memory. | 10 (7.5%) |
| Creative & Symbolic Expression | Edits intended to introduce artistic or metaphorical elements that represent an abstract feeling or idea associated with the memory. | 10 (7.5%) |
| Technical Improvements | Edits focused on enhancing the objective quality of the image. | 6 (4.5%) |

Table 4: Edit type taxonomy of participant prompts and frequencies

| Category | Definition | Example Prompts | Frequency |
| --- | --- | --- | --- |
| Modification | Change the existing elements. | "Change the boy's face to the smile face" <br> "Make the people in the photo smaller" | 62 (46.6%) |
| Addition | Create new elements into the photo. | "Add a modern two storey duplex model to the table" | 33 (24.8%) |
| Removal | Delete existing elements from the photo. | "Remove the people in the background" | 29 (21.8%) |
| Extension | Expand the original scene or introduce new content. | "Make the background complete" <br> "Extend the background" | 9 (6.8%) |

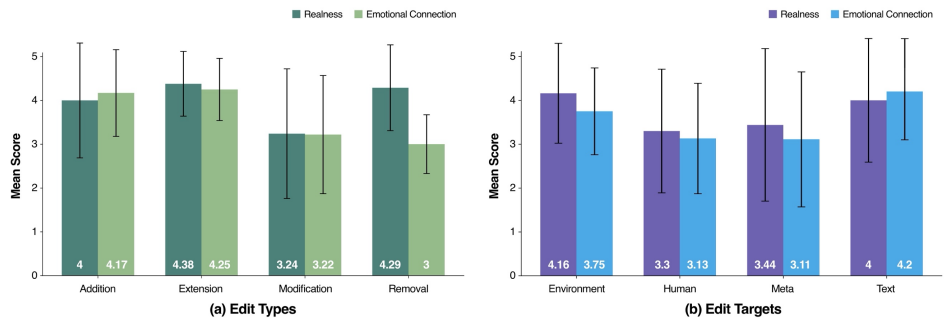

Figure 4: Mean ratings (±SD) of Realness and Emotional Connection. Sample sizes for (a): Addition (n = 29), Extension (n=8), Modification (n=52), Removal (n=28). Sample sizes for (b): Environment (n=57), Human (n=46), Meta (n=9), Text (n=5). Sub-category counts are reported in Appendix A.2.3

actions such as additions (n=33) and extensions (n=9) were less common, suggesting that while participants engaged with the generative potential of AI, their primary use was restorative and corrective. Across RX dimensions, State of Mind and Social Context most often led to Modifications (both n=21/62), suggesting that when people focus on feelings or social presentation, they tend to tweak existing content. Physical Context prompts produced a more balanced mix of edit types, while Memory Content prompts stood out for favouring Additions (n=13/33), indicating a stronger tendency to add new details when aligning the photo with what is remembered.

**Edit Target:** The analysis of edit targets shows that participants' edits around two primary elements: the environment (n=62) and the human subjects (n=54). Within environmental edits, the focus was often on specific objects (n=31). Edits targeting human subjects were most often directed at appearance (n=18) and the face (n=16), reflecting an engagement with identity and self-presentation in RX.

Table 5: Edit target taxonomy of participant and frequencies

| Category | Sub-category | Definition | Frequency | Sum |
|---|---|---|---|---|
| Environment | Object | The specific items within the photo (e.g., ballon, car). | 31 | 62 (46.6%) |
| | Natural | The natural elements (e.g., trees, river, mountains). | 10 | |
| | Artistic | The artistic or metaphorical elements (e.g., superman costume) | 9 | |
| | Whole | The whole environment. (e.g., background) | 5 | |
| | Weather | Weather conditions (e.g., sunny, rain, clouds). | 4 | |
| | Light | Explicit light-source. (e.g., sunshine) | 3 | |
| Human | Appearance | People's overall look, including clothing or accessories. | 18 | 54 (40.6%) |
| | Face | Facial features (e.g., expression, gaze). | 16 | |
| | Body | Body posture, shape or gesture. | 11 | |
| | Secondary | Other people in the photo who are not the central focus (e.g., crowds, passersby). | 6 | |
| | Core | Whole person (e.g., scale, reposition). | 3 | |
| Meta | Brightness | Overall lightness or darkness. | 4 | 11 (8.3%) |
| | Contrast | Overall difference between the light and dark areas. | 2 | |
| | Colour Temperature | Overall colour balance. | 2 | |
| | Size | Overall scaling or cropping of frame. | 2 | |
| | Clarity | Overall sharpness, focus, or resolution. | 1 | |
| Text | - | Textual elements (e.g., captions, dates, labels). | 6 | 6 (4.5%) |

### 4.1.3 Perceived Realness and Emotional Connection.
To examine how participants perceived different edits, we asked them to rate each edited photo on two 5-point Likert scales: Realness and Emotional Connection. Detailed descriptive statistics are reported in Figure 4.

Across edit types (Figure 4a), Extensions (realness M=4.38, SD=0.74) and Removals (M=4.29, SD=0.98) tended to be rated as more realistic, whereas Modifications received the lowest realness scores (M=3.24, SD=1.48). Within this dataset, the average realness rating for Removals was about one point higher than for Modifications on the 5-point scale. For emotional connection, Extensions (M=4.25, SD=0.71) and Additions (M=4.17, SD=0.99) tended to evoke stronger emotional connection than Removals (M=3.00, SD=0.67), with Modifications falling in between. Taken together, these descriptive differences suggest that while Removals best preserve perceived realness, Extensions and Additions more often enhance emotional connection.

A similar pattern appears across edit targets (Figure 4b). Environment edits received the highest realness ratings (M=4.16, SD=1.14), whereas Human edits were rated lowest (M=3.30, SD=1.41), particularly those involving faces (M=2.80, SD=1.47). In our sample, Environment edits were on average about 0.9 points higher in realness than Human edits. For emotional connection, the differences between targets were smaller: Environment edits tended to elicit somewhat higher emotional connection (M=3.75, SD=0.99) than Human edits (M=3.13, SD=1.26), while Text edits, though rare (n=5), were rated highly (M=4.20, SD=1.10). Overall, these patterns suggest that participants were more comfortable—both perceptually and emotionally—modifying environmental elements than human subjects.

### 4.1.4 The Relationship between Edit Target and Edit Type.
To understand the relationship between edit target and edit type, we analysed the flow from the four target categories to the specific edit types performed (Figure 5).

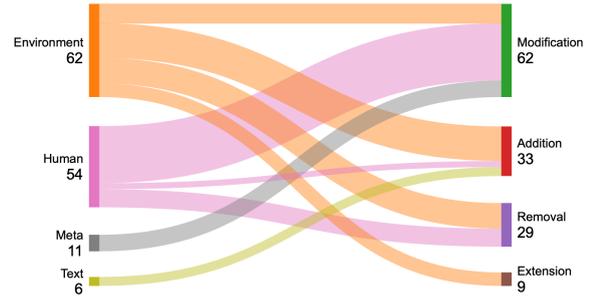

Figure 5: Flow from Edit Targets to Edit Types. The diagram illustrates how participants' choices of what to edit (left) influenced the kind of action they performed (right). The thickness of each flow is proportional to the number of edits for that path.

Most human-targeted edits were Modifications (n=38), primarily aimed at refining appearance or facial features. Generative actions like Additions (n=4) were rare. This indicates that when engaging with core identity elements, participants adopted a cautious, restorative approach, preferring to adjust existing features rather than create new ones or make drastic changes.

In contrast, edits targeting the Environment were more varied and generative. While Modification was still a common strategy for environmental elements (n=13), participants were far more likely to perform generative actions like Additions (n=23) and Extensions (n=9) in this domain. This suggests that participants viewed the environment as a flexible and expandable canvas, using the GenAI's creative capabilities to add new objects, alter the atmosphere, or expand the scene to better align with their memory.

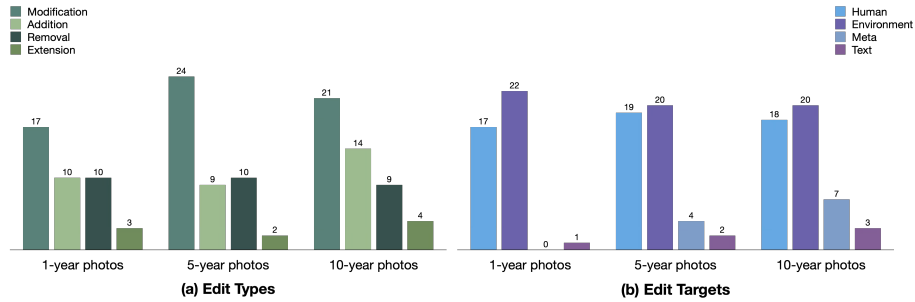

Figure 6: Distribution of Edit Types (a) and Edit Targets (b) across photo ages.

Overall, participants applied a wider variety of edit types to environments, treating them as flexible and expandable, whereas edits to humans were more narrowly focused and restorative. Edits to other targets, such as text or meta attributes, followed patterns shaped by their functional properties.

## 4.2 The Influence of Photo Age on Editing

To investigate how the age of a photo and the memory it represents influence editing practices, we analysed the distribution of both Edit Types and Edit Targets across photo age (1, 5, and 10 years), as shown in Figure 6. Excluding skipped tasks, the number of edit requests increased with photo age: 1-year photos generated 40, 5-year photos 45, and 10-year photos 48 edit requests. In addition, 39 out of 144 RX Tasks (27%) were skipped. Skips were spread relatively evenly across photo ages (1-year: 14, 5-year: 12, 10-year: 13), suggesting that photo age alone did not strongly determine whether participants chose to edit or not.

For edit types (Figure 6a), Modifications dominated at all photo age groups (n=17, 24, 21). Generative actions were more frequent for older photos: Additions peaked at 10-year n=14 (vs. 1-year n=10, 5-year n=9), and Extensions were also higher at 10-year n=4 (vs. n=3, 2). Removals were relatively stable (n=10, 10, 9). For Edit Targets (Figure 6b), Environment edits were consistently highest and stable (n=22, 20, 20), and Human edits were steady (n=17, 19, 18). Meta edits rose with age (1-year n=0, 5-year n=4, 10-year n=7), and Text edits, though rare, also increased (1, 2, 3).

Overall, these descriptive patterns suggest that older photos attracted slightly more generative and restorative edits, while attention to people and environment remained comparably steady.

## 4.3 Beyond the Edit: How GenAI Shapes RX

We conducted a reflexive thematic analysis of 12 post-editing interviews, which ranged from 37 to 57 minutes in length (M = 44), to investigate how participants experienced editing and viewing GenAI-edited photos. We identified five themes that capture the broader motivations and impacts of editing. In presenting these themes, we integrate illustrative participant edits with earlier quantitative findings to provide triangulated insights.

### 4.3.1 Aligning Photographic Records with Felt Memory.
Our analysis shows that participants' primary motivation was not factual accuracy but closing the gap between the photo's visual record and their *felt memory*. For participants, what mattered was not strict factual representation but whether the edit aligned with how the moment was remembered and experienced.

**Mental impression over realness:** Participants (n=12) consistently articulated a preference for mental impression over realness, using the tool to make edits that would *"fit the mood"* (P11M) and *"I'm happy even though it's not real"* (P8F). P7F described their process as an effort to *"edit to align the memory in the imagination,"* explicitly framing the goal as a calibration of the external artifact to match an internal, idealised conception of the past. Sometimes alignment was literal: if a detail in the photo did not match what they remembered, they changed it. For example, P8F remembered a day being unbearably hot even though the photo looked overcast; using the GenAI, she added sunshine in the photo, *"edited to restore the hot weather"* to reflect their lived experience. Others (n=5) added elements that they felt were part of the memory but missing from the picture, effectively using GenAI to *bridge gaps* between the photographic record and their remembrance. These cases correspond to the "Memory Alignment" edits observed in the intention analysis.

**Aligning photos with identity:** In aligning photos to their personal memories, participants frequently adjusted depictions of people, themselves especially, to match personal standards or feelings about the past. Several admitted to subtle appearance adjustments, like changing attire or facial expression, so that the photo would concord with how they wanted to remember that moment. For example, P11M removed an object from his own image *"to look healthier"* in retrospect. P6M talked about editing his clothing and posture in the photo to better fit the nervous mood he remembered having. These edits reflect an interplay of social context and state of mind. The result was a more personally satisfying memory artifact: the edited photo felt truer to them if not strictly true to reality.

**Curating the memory for sharing:** In addition, participants editing photos not only conforms to their own 'imagination,' but also allows others to better understand their memories. P1F make an interesting change in her 5-year photo (Figure 7). The happy birthday balloons behind her were a bit messy, so she wanted to use GenAI to arrange them neatly. She explained that she wants to

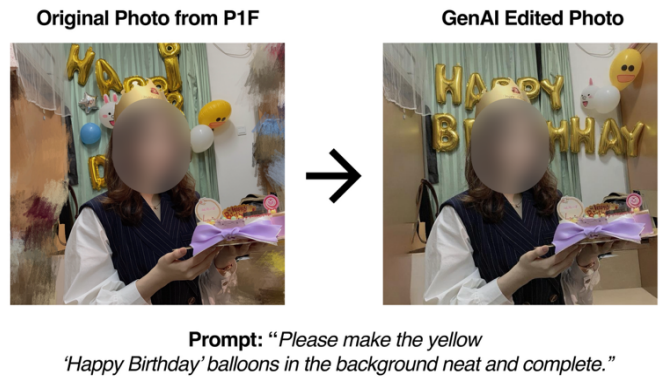

**Figure 7: P1F's prompt to arrange the 'Happy Birthday' balloons resulted in an edited photo where 'Birthday' was misspelled.** *Note:* The photos have been cropped to emphasise the details.

*"make it easier for others to understand"*. When she saw the edited photo, even with the misspelled 'Happy Birthday' balloons and other not prompted but changed background details, she said: *"I actually like this version. I feel [edited photo] is more in line with the scene I imagined, the scene in my memory. Because in my memory, when the original photo was taken, some of the 'Happy Birthday' letters had fallen off. But when I open the door and first saw those ballons, they were like the edited one. So [edited photo] is better matches the scene in my memory."* She rated this version a 4 out of 5 for realness.

This theme is supported by our thematic analysis of intentions, where "Emotional Amplification", edits intended to enhance the mood or feeling, was the single most frequent intention for editing. This pursuit of emotional remembering experience creates a fascinating paradox in how participants evaluated the edits. Edits that were factually inaccurate, such as adding sunlight to a cloudy day, were rated as *"more real"* than the original photograph (P4M). For example, edits targeting environmental elements to enhance mood, such as "Weather" changes, received high mean ratings on both emotional connection (M=4.75, SD=0.50) and perceived realness (M=4.75, SD=0.50). Participants' benchmark for "realness" was not the objective conditions captured by the camera, but their felt memory of the moment.

*4.3.2 Acceptable and Unacceptable Edits: People Fixed, Context Matters.* Most participants (n=9) established a clear hierarchy of what they considered editable versus inviolable. Edits that touched on a person's identity, such as faces and bodies, were considered unacceptable, as were edits involving important objects that served as memory cues. By contrast, edits to the environment were broadly accepted. Participants often modified, expanded, or added environmental details to support reflection, sharing, and storytelling around the memory.

**Faces as anchors:** This boundary was consistently articulated in participant interviews: the person in the photo was treated as a non-negotiable anchor, when changes related to face. As P5F explained, altering the face made the photo feel like *"looking at another person."* Such reactions reflect a broader discomfort with identity-related changes, especially since GenAI edits to human faces were often inconsistent and unpredictable. Two participants, however, reported exceptions where the generated expressions closely matched their own, leading to positive evaluations (P7F, P8F). P8F changed the smile and said: *"I still like the smiling."* P7F described the edited 10-year photo as *"actually very beautiful … I feel the smile is very good,"* but in contrast found the 5-year photo *"doesn't look like me. It feels uncomfortable."* Overall, edits targeting Human (Face) were the most poorly rated category. They received a mean realness score of 2.80 (SD = 1.47), falling below the neutral baseline of 3 and indicating negative responses to facial alterations.

**Conditional acceptance of appearance and body edits:** In contrast to the strong resistance toward facial edits, participants (n=6) were generally more accepting of GenAI edits to appearance (e.g., clothing) and body posture (e.g., gestures). Acceptance, however, was closely tied to whether the GenAI could execute edits in a way that felt consistent and aligned with the user's intent. For example, P8F used GenAI to add a Hawaiian shirt to a beach photo of her partner, noting that he owned a similar shirt and describing the result as *"very natural."* P7F also changed her own top from black to white to appear more energetic and expressed satisfaction with the outcome. Gesture changes were similarly well-received: P1F and P9M both replaced hand positions with a "V-sign" and rated the results positively. At the same time, limitations in generative precision were evident. P11M requested that the GenAI take his hand out of his pocket in a 10-year photo; while he felt the output was convincing *"at a glance,"* closer inspection revealed hands that felt subtly *"different from my own."* Quantitative ratings reinforce these perceptions: edits to Human (Body) and Human (Appearance) achieved mean realness scores of 3.43 and 3.29, respectively, indicating cautious approval of GenAI's capacity to produce plausible results, provided that execution quality was stable and coherent.

**Environment extension as safe space:** Participants established clear boundaries for environmental edits, welcoming changes that enriched a scene but rejecting those that altered its fundamental story. They were comfortable with edits that extended or clarified a background, as these changes preserved the memory's core narrative while making the photo more vivid. Quantitative findings support this distinction: Extension edits were exclusively applied to the environment (9/9 instances) and achieved the highest mean ratings for both realness (M = 4.38, SD = 0.74) and emotional connection (M = 4.25, SD = 0.71).

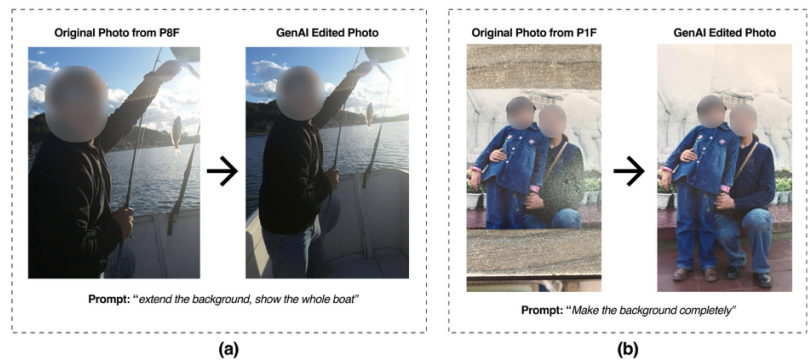

Figure 8: Examples of environment-focused GenAI extensions. On the left (a) P8F's fishing trip photo, extended to reveal the entire boat. On the right (b) P1F's childhood photo with her mother, extended to reconstruct missing background context.

For example, P8F used GenAI to expand a constrained view of a fishing trip photo to show the entire boat (Figure 8a). Although the edit shifted the viewpoint from a side angle to the bow, she was highly satisfied: *"It's a better angle to show than the original"* She further reflected on its personal significance: *"It's important because to know I was on a boat. If time goes on, I might forget what the boat looked like."* In this way, extensions aligned with the remembered experience, helping her preserve contextual details that might otherwise fade. Similarly, P1F edited a 10-year photo with her mother, where only a mobile copy of the original remained (Figure 8b). She explained that seeing more of the background evoked additional memories: *"It didn't tamper with the information in my original photo. It just enriched it based on the information in the original photo."* Such edits were perceived as acceptable because they remained within the plausible bounds of the original scene, enriching the remembering experience without overwriting its content.

**Cultural mismatches:** In addition to technical inconsistencies, several participants also reported culturally mismatched generations that further reduced the acceptability of identity-related edits (n=4). These mismatches appeared even when the requested change was simple. For example, P3F asked the tool to open a car door; although the original photo showed a right-hand-drive vehicle common in Australia, the model generated a left-hand-drive layout. Asian participants (P1F, P2M, P5F) similarly expressed discomfort when their smiles were transformed into broad, white-toothed expressions—an aesthetic aligned with Western portrait norms. As P2M noted, *"It doesn't feel like my own smile."* When GenAI outputs conflicted with participants' cultural expectations or lived context, dissatisfaction increased and the edits were judged as unacceptable.

In summary, the participants goal was not just to improve a photo aesthetically but to protect the narrative integrity of the memory itself. The face of a loved one or a significant object acts as a primary memory cue, a potent trigger that grounds the connection between the artifact and the remembered experience. Altering this anchor threatens to sever that link, creating a *different* story altogether. Modifying the periphery, however, can strengthen the focus on the primary cue, as exemplified by the "Distractions Refinement" intention, which accounted for 17 edits by 8 participants aimed at removing irrelevant elements.

*4.3.3 Reconstructing the Remembered Experience with GenAI Edits.* Participants used the GenAI's capabilities not only to refine existing content but also to actively co-construct and augment their remembering experiences.

**Re-experiencing through edits**: Participants (P4M, P5F, P7F) explicitly likened the experience of viewing edited photos to re-experiencing the original scene. All reported that editing and seeing the edited photos was like re-experience the memory from back then. P5F reflected: *"After the modification, I can feel… when I saw the river channel at that time, I was captivated by its scenery. I felt, 'Wow, I have to capture this.' I re-experienced the mood I was in when I took the photos"*. Similarly, P4M noted: *"The process of editing all the photos is like re-experiencing the whole thing."*

**Edits to add memory cues:** In our study, although some edit requests target at Meta (e.g., brightness, contrast, colour) and Text were unsuccessful (two failed Meta adjustments; one failed text addition, and two instances of text misspelling), participants nonetheless demonstrated the value of these modifications as memory cues. Several (n=4) described using GenAI Meta edits to repair faded photos, making the artefact clearer and closer to how they remembered it. Similarly, four participants used text additions to annotate location, people, and temporal information as supports for remembering. For example, P2M added text labels to both a 10-year photo and a 5-year photo, representing a school trip with classmates and a university graduation ceremony, respectively. He explained: *"Remembering a specific date is meaningful to me; I can use that date to recall more about the event."*

**Editing process as reconstructing**: Editing process itself became a valuable reconstructive remembering practice. All participants (n=12) reported noticing new details or associations during editing, often linking the photos to previously forgotten events.

P2M observed: *"After the modifications, I can remember the scene of us all playing together."* P3F remarked that *"editing the photo... really did trigger more of my memories."* P12F similarly described how the process *"makes me... remember this event more deeply. It deepens my memory."* These accounts are consistent with findings from other participants (n=6) who indicated that edited photos drew their attention to some overlooked aspects, such as clothing, posture, or background elements, that in turn helped to reconstruct forgotten details.

This theme illustrated the active, situated role GenAI can play in shaping participants' recollections. The interaction with GenAI was itself a remembering activity, not merely image production. Thus, GenAI editing enriched and transformed participants' RX, shifting it from simple reminiscence toward dynamic, creative reconstruction: memories were expanded, reinterpreted, and actively re-built.

*4.3.4 Memory as Canvas: Creative Expression.* Beyond reconstructing, participants (P3F, P10M, P12F) also embraced GenAI as a means of creative reimagining, treating the original memory as a foundation upon which to build new, symbolic narratives. This theme highlights remembering as an imaginative, interpretive process: some edits were about adding symbolic or fantastical elements to evoke meanings and emotions associated with the memory. Participants particularly engaged in such edits under the state of mind and social context prompt, when they reflected on the broader significance of the memory.

**Creative augmentation:** For example, P12F experimented with stylistic filters and whimsical additions to transform a photo into a more artistic representation of her experience: *"not to change what happened, but to express it better."* Similarly, P3F introduced artistic elements to express deeper emotions and significance embedded in their memories. P3F revisited a photo from a trip arranged by a friend a decade earlier, which captured a moment just before attending a concert during a turning point in her life (Figure 9). She described the small journey as pivotal in helping her recover from difficult experiences and transition into a new phase of independence: *"It was with the help and support of my friend that I felt... like a person is like a tree, slowly growing my own... clear ability to comfort myself. It felt like I started to take root from that moment."* To express this, she asked the GenAI to add a dark, root-like structure with a white flower to the photo, symbolising growth emerging from hardship. Although the generated image did not fully match her expectations, she was *"surprised"* at its realism and appreciated its mnemonic value. She explained that the symbolic element helped make fragmented memories more concrete: *"With this edited photo, perhaps because it was edited by my own command, I can see this abstract object at a glance and think, 'Oh, it's because I had this feeling at the time.' So it has a bit of an assistive role, making fragmented things more concrete."*

**Contextual symbolism for life events:** P10M reflected on a five-year-old photo documenting a decisive moment when he and his wife chose to invest in building a new house (Figure 10). He requested GenAI to insert two architectural models under the *social context* dimension. He described being impressed by the result: *"It's quite impressive the AI figured this out."* For him, the models not only anchored the moment but also held potential future value: *"Maybe when I get older and I suffer from dementia or whatever... the*

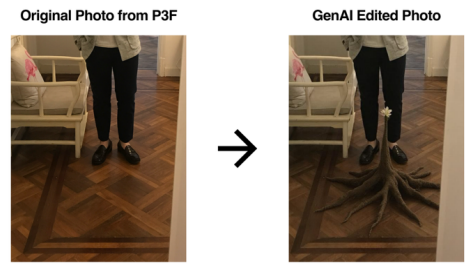

**Figure 9: Symbolic augmentation through GenAI editing.** P3F added a root-and-flower to represent personal growth and emotional recovery during a pivotal life stage. *Note:* The photos have been cropped to emphasise the details.

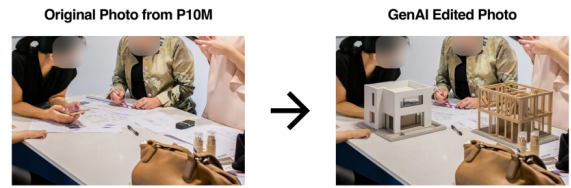

**Figure 10: Representing life decisions with generative cues.** P10M inserted duplex models into a photo to symbolise a new home-building decision and to serve as a mnemonic anchor for future and intergenerational remembering. *Note:* The photos have been cropped to emphasise the details.

*model could serve as an anchor to remind."* He further emphasised the intergenerational role of photos: *"You want to use the photos to educate the next generation as well, to let them know what the current and previous generation have done. I'm trying to document that process. I think to have some realistic model would help."*

Quantitative findings reinforce this theme: Environment (Artistic) edits received high emotional connection ratings (M = 4.22, SD = 0.83), second only to Environment (Weather). Overall, GenAI's abilities allowed participants to engage with their past in a richly interpretive way. This theme underscores the potential for generative technology to support remembering as a creative, meaning-making process. GenAI editing opened a space for participants to express themselves, infusing personal symbolism and creativity into their remembering experience.

*4.3.5 Risks and Concerns in GenAI Photo Editing.* **Unsuccessful edits:** Across all editing requests, 12.0% (n=16) were unsuccessful edits. In Appendix A.3 we provide a detailed breakdown of unsuccessful edit types and targets. In summary, such unreliability, particularly in human representation, led to frustration and disrupted the collaborative process. Several participants (n=6) noted that *"the GenAI didn't follow the prompt strictly"* or that *"people edit [was] lacking"* when instructions were misinterpreted or inconsistently applied. Notably, edits targeting humans accounted for 8 of the unsuccessful edits. In practice, this technical unreliability shaped how participants approached later edits. Several participants (n=5) responded to repeated failures by making their prompts more fine-grained, or by explicitly stating that the rest of the photo should remain unchanged. Others (n=4) described "testing" the model with small changes before attempting anything meaningful, indicating a cautious, probe-and-adjust strategy.

**Undesired loss of contextual cues:** More than half of participants (n=7) noticed loss of contextual information during the editing process. P7F explained that when she removed background crowds, the pier, a distinctive feature that had originally anchored her memory of the location, also disappeared, leaving her less able to recall the exact place. Similarly, P6M observed that some information loss felt subtle and almost unavoidable, yet this raised unease: even small missing details left him *"feel[ing] insecure about this situation."* These reflections suggest that generative editing may inadvertently strip away memory anchors, destabilising the sense of place and detail that supports remembering.

**Potential Memory Change:** Participants (n=8) expressed concern that the more convincing GenAI's outputs appeared, the greater the risk of potentially changing their personal memories. As P1F noted, *"[after the editing] I feel like my memory has been altered"*. P7F described how during a trip she had swapped hats with a friend and was photographed wearing the friend's hat. In the study, she asked the GenAI to change it with her own hat to *"look more energetic."* However, in the interview she reflected: *"The edited version looked so real that if it stayed on my phone, years later I might forget the little story about swapping hats with my friend."* P8F also worried about long-term effects, reflecting, *"Maybe after ten years, when I see that [edited] face, I can't say exactly if my memory is still the same or not"*.

**Concerns about Misuse:** Participants (n=5) also raised concerns about broader implications of manipulation and misuse. P10M described the edits as a *"double-edged sword,"* pointing to the risks if *"bad people want to use it to manipulate people"*. P9M extended this unease to legality and ethics, questioning whether sharing GenAI edited photos online might contravene rules: *"If I'm sharing the [edited] car plate online, am I against the regulations?"*.

Together, these themes illustrate the dual role of GenAI photo editing in remembering practices. On one hand, GenAI empowered participants to align photos with felt memory, enrich or creatively reinterpret memories, and reconstruct remembering experience. On the other, its unreliability, unintentionally to erase contextual cues posed notable risks. The remembering experience emerged as a negotiated space where participants balanced enrichment against distortion, selectively adopting edits that supported their sense of memory while resisting those that undermined it.

## 5 Discussion

Our findings suggest that GenAI is reshaping purposive "PhotoUse" [6]. Participants used GenAI to actively reconstruct memories in service of specific goals. In this discussion, we first conceptualise this practice as constructive PhotoUse guided by felt memory (Section 5.1) and outline the hierarchy of editability users employ to preserve authenticity (Section 5.2). We then discuss the potential benefits (Section 5.3) and risks (Section 5.4) of these capabilities, before proposing design commitments for responsible GenAI (Section 5.5).

### 5.1 The Primacy of Felt Memory

Our results show that when participants used GenAI to edit personal photographs, their primary motivation was not to produce a precise record of the past, but to reconcile discrepancies between a photo's visual content and their felt memory of the event. As mentioned in the Introduction, we use felt memory as a shorthand to emphasise the role of subjective experience in remembering [74, 81], highlighting how participants sought resonance and meaning rather than factual accuracy.

**From curation to construction.** This observation empirically validates the shift in PhotoWork identified in Section 2.1. Prior research by Kirk et al. [31] and Petrelli et al. [61] conceptualised PhotoWork primarily as curation, defined as the selecting, arranging, and managing of existing memory media to support recall. Our study demonstrates how GenAI actualises an emerging mode we term *constructive* PhotoUse, in line with PhotoUse [6]. In this mode, users do not merely choose among available cues but actively synthesise and modify them using the semantic editing affordances. For example, rather than making corrective edits to repair the image, P3F actively reconstructed the scene by adding a symbolic root and flower to reinterpret the memory through the lens of later growth. Consequently, GenAI broadens the scope of PhotoUse, expanding it from a practice of preservation to one that also encompasses collaborative reconstruction.

**Malleability for feeling, not deception.** This perspective aligns with psychological accounts of memory as a constructive process [2, 70] but challenges the dominant discourse on GenAI in Section 2.3. While existing literature often frames the GenAI edited media as a risk factor for misinformation and false memory [59], our findings suggest that in the context of personal remembering, malleability serves a different function. By structuring prompts through the RX dimensions, participants re-created the past in light of present purposes, with GenAI serving as a mediator in this reconstructive process. Thus, factual inaccuracy is not necessarily a failure of the system, but often an intentional alignment with felt memory.

### 5.2 Hierarchy of Editability in Personal Photos

This hierarchy, which prioritises the preservation of identity anchors while permitting environmental modification, reveals how users navigate the tension between photographs as prosthetic memory [13, 42] and as malleable media. While GenAI technically affords the seamless alteration of any pixel, our participants imposed a strict, self-regulated boundary between identity anchors and contextual elements.

**Protecting the anchor.** Participants treated human features, particularly faces, as inviolable. As P5F explained, altering the face made the photo feel like *"looking at another person."* This resistance aligns with the psychological role of faces in anchoring self-identity [75], but also serves as a defence against the homogenising effects described in Section 2.3 [11]. Participants consciously rejected edits to human subjects avoiding the risk to replace unique identity features with culturally homogenised defaults.

**Generative autotopography.** Conversely, the environment emerged as an active, malleable resource, extending the HCI concept of "autotopography" [62, 63]. Originally, autotopography described how people arrange physical mementos in their homes to support memory. We extend this concept into the generative realm. In our study, participants actively "arranged" digital environments such as adding sunshine, removing crowds, or extending backgrounds to create a new scene that better supported their felt memory. Unlike traditional autotopography which is constrained by physical space, generative autotopography enables users to virtually reshape the spatial context in a more fluid way. Thus, contextual edits were not considered as falsifications, but as exploratory enhancements to support vivid re-experiencing [23, 45, 52].

### 5.3 Positive Impact and Applications of GenAI Photo Editing

Participants recognised benefits for remembering when GenAI is used deliberately. We outline three application areas: everyday reminiscence, family storytelling, and health and care.

*5.3.1 Personal Wellbeing and Everyday Reminiscence.* Month-long smartphone-photography interventions can improve affect [33], and end-of-day picture review can strengthen autobiographical memory [20], indicating that daily, low-effort reflection on personal photos can be beneficial. In our study, first-person, self-authored GenAI editing showed similar potential (see 4.3.3–4.3.4). We envision a guided-reflection mode in mainstream photo apps: brief prompts (e.g., "How did you feel that day?") followed by context-first suggestions (light, weather, tidying) with instant preview and one-tap revert. This turns editing into low-effort journaling that supports mood and reflective review, extending practice from capture and revisit to editing for felt memory.

*5.3.2 Family Archives and Intergenerational Storytelling.* GenAI can strengthen family archives by clarifying narratives and increasing mnemonic cues. Family memories are tightly coupled to situated context and shared stories, while domestic digital collections often lack such scaffolding and impose heavy *PhotoWork* [62]; "technology heirlooms" research highlights challenges of provenance, meaning preservation, and narratability over time [53]. GenAI offers a practical path to restore contextual cues and lighten curation via straightforward environmental adjustments and lightweight text (e.g., dates, locations). Co-editing can itself become a storytelling occasion where families negotiate what to keep or change and co-construct narratives—turning intergenerational review into a scaffolded, pass-downable practice.

*5.3.3 Health and Care Settings.* In our study, eight participants independently proposed GenAI editing to assist people with memory difficulties. Evidence shows reminiscence therapy can improve quality of life, mood, communication, and engagement, with digital delivery improving depression and cognition relative to treatment as usual [47, 64, 80]; HCI co-design points to needs for accessible tools that preserve cues and support lightweight storytelling and annotation [83]. GenAI should prioritise low-risk edits (restoration, clarity, colourising, tidying, cue-adding) and support caregiver-led narration. Edits to faces or bodies warrant extra caution; emphasise subtle environmental enhancements aligned with cue-preserving design principles [27].

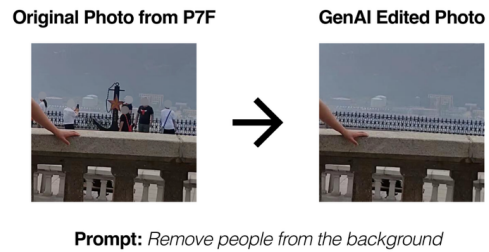

**Figure 11: Context loss during an edit. The landmark pier emblem was unintentionally removed along with bystanders.** *Note: The photos have been cropped to emphasise the details.*

### 5.4 Negative Impact of GenAI Photo Editing

While Section 5.3 outlined potential benefits and applications, this section addresses the accompanying risks. Even when used carefully, GenAI editing can cause information loss, impose Western-centric defaults that diminish cultural diversity, and raise ethical and privacy concerns.

*5.4.1 Information Loss During Editing.* Even when GenAI produce visually consistent results, our participants indicated that the edited photos almost inevitably lose information. At the "minor" end, some edits routinely induce colour shifts [58, 65]. At the "major" end, remove operations can delete unintended content beyond the prompt target [77]. When P7F tried to *"remove people"*, the GenAI also deleted a pier landmark, collapsing a memory cue (see Figure 11). In the interview she reflected, *"After that landmark was removed, for a moment I asked myself whether this was [location] or [location]."* As GenAI diffuse into everyday photo apps, the cumulative effect of such micro-losses is likely to grow. When provenance is absent and edited versions overwrite or eclipse originals, users face source confusion: the edited photo becomes the most accessible "record."

Studies underscore why these dynamics matter. Exposure to doctored photographs can create or inflate confidence in false memories [21, 76], especially after AI edited [59]. Even when participants (n=4) in our study judged some lost details as "unimportant" for their own remembering, the same losses can be consequential in forensic, news, or clinically vulnerable contexts, where small cues support orientation and trust. Recent public controversies over edited official photos further illustrate how minor fixes can undermine public confidence without robust provenance [18].

*5.4.2 Cultural Bias in Models.* Text-to-image and editing models often embed Western-centric defaults that surface even in seemingly neutral operations. Prior audits of diffusion models such as DALL·E 2 [84] and Stable Diffusion [97] consistently report representational skews that align with U.S. demographics and aesthetics [3, 50]. These biases are traceable in part to training corpora with uneven geo-diversity, where widely used image datasets over-represent North America and Europe and under-represent other regions, producing performance and depiction differences across locales. Recent studies quantify cultural representativeness and show that generations for places and subjects frequently drift toward Western visual tropes even when prompts specify other cultural clusters [41, 82].

In our study, these dynamics shaped how some participants, particularly Asian Australians, experienced identity-related edits. As prior work shows, smiling conventions differ across cultures and certain facial expressions that align with Western portrait norms are not universally appropriate [37]. When the model substituted participants' natural expressions with Western-styled smiles, the results conflicted with their sense of self, reducing acceptance and trust. Likewise, the model's assumption of a left-hand-drive vehicle reveals how Western defaults can overwrite contextually accurate regional details.

These patterns highlight important implications for generalisability. Although our sample was based in Australia, it included participants with diverse cultural backgrounds. This suggests that such biases may be amplified in regions less represented in training data [19]. Consequently, GenAI may risk narrowing the expressive range available to users by pulling personal memories toward culturally dominant visual norms. For memory-oriented applications, where authenticity and identity are foundational, this has practical and ethical consequences: models must be designed to support local conventions rather than overwrite them.

*5.4.3 Ethical and Privacy Concerns.* In HCI, responsible or human-centered AI is often framed as designing systems that reflect diverse human values and provide meaningful user oversight, while anticipating broader social impacts [34, 73]. GenAI-mediated remembering fits this framing, but shifts its centre of gravity: what matters is not only whether an edit aligns with a user's values today, but whether its provenance stays legible and socially accountable, when the image resurfaces months or years later. This makes responsibility longitudinal: edited photos can become the default record, so systems should preserve a clear, interpretable account of what changed and why, for the future self and for others. The increasing integration of GenAI into everyday tools necessitates a critical examination of privacy risks, the potential for misuse, and the subtle ways these systems influence our relationship with the past. A primary ethical concern in GenAI photo editing is the use of personal data. Many commercial services incorporate user-uploaded images into training datasets, often without sufficient transparency about how these data are stored, processed, or shared. This creates privacy risks, since personal photographs frequently contain sensitive or identifying information that could be inadvertently exposed or used to construct detailed individual profiles. Recent work shows that large generative models can unintentionally memorise and reproduce training data, raising the possibility of re-identifying individuals from supposedly anonymised datasets [7]. Investigative reports further highlight that even casual uploads to image generators may carry metadata such as GPS or device details, which can persist as hidden privacy risks [55]. To mitigate these risks, we deliberately selected the Flux.1-pro/kontext model accessed via the fai.ai platform. Unlike many commercial systems, this platform guarantees that participant photos are not used for retraining and are permanently deleted after each session. This ensured that all photo data remained confidential and under researcher control, aligning our methodology with emerging practices in privacy-preserving AI [38].

Beyond data privacy, broader ethical and societal risks have also emerged. Rini and Cohen [66] describes how deepfakes can cause "deep harms," including non-consensual sexual content, reputational damage, and even the erosion of a person's sense of self when their likeness is manipulated. At the same time, research highlight how GenAI enables new forms of digital deception in phishing and social engineering, where realistic synthetic media can manipulate trust and extract sensitive information [71]. Prior work on responsible AI for synthetic media has largely emphasised transparency interventions such as disclosure, labelling, and provenance to counter deception and protect trust [14]. Our findings extend this lens to everyday remembering, showing how self-authored edits can support felt memory while participants still protect identity anchors, which reframes responsibility as enabling expressive reconstruction alongside safeguards for cue preservation and source clarity. Together, these studies show that the risks of GenAI extend beyond individual misuse to systemic threats for memory, identity, and social trust. Exploring these questions is not just a technical challenge but a critical step toward fostering a healthy and responsible relationship with our digital pasts.

## 5.5 Design Implications

Our results suggest that GenAI can be designed as partners in remembering, rather than as neutral fixers of pixels. GenAI can respect identity boundaries, preserve orienting cues and make the work of memory visible, deliberate and shareable. Below, we summarise the implications in the form of a coherent set of design commitments with concrete consequences for interaction, feature scope, and evaluation.

*5.5.1 From Editor to Remembering Partner.* Participants' edits were less about repairing defects than about aligning a photograph with how the moment is now remembered, including its atmosphere, meaning, and narrative role. Tools can therefore centre co-narration and reconstructive work over correction. These tools can provide goal-aligned modes (for example, Social Curation versus Memory Reconstruction) that surface different defaults and prompts tied to RX goals. In reconstruction mode, systems can foreground ambient adjustments and narrative cues over pixel-perfect retouching. Meaning scaffolds can also be integrated at edit time, such as lightweight text or voice annotations prompted after environmental changes (for example, "You added rain; what sensation do you remember?"). This way, reflections remain attached to the edit and can travel with the photo when desired. These shifts align the system with participants stated preference to "fit the mood" rather than factual accuracy.

*5.5.2 Designing for Mindful Memory Work.* Edits can reshape later recall, particularly when edited versions overwrite originals or when micro-losses accumulate. The interface can therefore encourage mindful and transparent changes. A persistent edit history line showing the relationship between the original and the edited image, combined with per-edit visual comparisons and easy non-destructive rollback, can reduce source confusion. Temporal friction should be introduced for high-impact actions on older images, such as adding or removing people, since our data show that transformative edits cluster on 10-year photos. A short reflective confirmation such as "This is a 10-year memory, do you want to proceed?" can be sufficient. Finally, machine-readable provenance should be embedded when exporting so that edited images remain auditable as they circulate across time and contexts.

*5.5.3 Context-first, Identity-safe Defaults.* Participants clearly differentiated identity anchors (faces, bodies) from flexible environments. Interfaces should enforce identity protection by default, foregrounding environmental adjustments (lighting, background cleanup, scene extension). Editing human features should require explicit, reversible overrides with risk-level indicators and "still looks like me?" self-checks. GenAI can warn about edits impacting orientation cues (e.g., landmarks, signage) and propose safer alternatives. Additionally, generative models must respect cultural contexts, offering context-aware toggles to maintain scene authenticity.

*5.5.4 Guardrails Against Information and Cue Loss.* Even convincing generations can erase orientation cues (e.g., landmarks) or introduce subtle drifts (e.g., colour shifts), undermining trust and remembering. Tools should include cue-preserving features: a protect-cues brush for signage, landmarks, and timestamps; pre-removal warnings when edits intersect likely cues; and safer alternatives such as blurring rather than deletion. Linking these guardrails to the provenance layer ensures users see what changed and why. Such measures respond directly to participants' reports of unintended deletions and the risk of source confusion when edited images eclipse originals.

## 5.6 Limitations and Future Work

This study provides an exploration of how participants use GenAI to edit personal photographs and the impact of editing process and edited photos on the remembering experience. However, our study has several limitations, which in turn open up new avenues for future research.

A primary limitation of our work is the small, culturally homogeneous sample of 12 participants based in Australia, which restricts the generalisability of our findings. The reliance on qualitative analysis also limits the statistical power to make broader claims. Future research should therefore recruit larger, more culturally diverse participant pools to validate our findings through a mixed-methods approach, which could reveal how cultural values shape photo editing motivations.

Furthermore, the study was conducted in a university setting using provided equipment. While this ensured systematic data collection, it may not reflect how people naturally interact with GenAI in their personal, everyday contexts. To capture more authentic behaviours, future work should employ "in the wild" methods like diary studies [8]. Crucially, longitudinal studies are needed to understand the long-term effects of these practices, such as whether edited photos enhance, alter, or even replace original memories over time.

Finally, our findings are inherently linked to the specific GenAI tool used, whose affordances and interface shaped user interactions. To distinguish our findings from the influence of a single technology, future research should conduct comparative analyses across different editing platforms (e.g., text-prompt-based versus direct manipulation interfaces). This would clarify how specific design choices, such as text-prompt versus direct manipulation interfaces, influence the editing process and user experience.

## 6 Conclusion

Our study examined how GenAI photo editing reshapes the remembering experience. Through a two-phase study with 12 participants, we showed that people privileged felt memory over factual accuracy, guided by a clear hierarchy of editability that protects human elements while treating environments as flexible scaffolds. Editing itself became a form of active reminiscence: participants not only edited photos but also re-experienced, curated, and even creatively reinterpreted their pasts. Yet this co-creative potential was tempered by risk. Our participants encountered unsuccessful edits and unintentional information loss. They also surfaced concerns about cultural bias and broader misuse, highlighting that technologies designed for personal meaning-making can still reproduce systemic limitations. These tensions underscore the double-edged nature of GenAI as both a memory partner and a destabilising force. We provide empirical accounts of how people use GenAI to edit personal photos for remembering, showing how intentions, edit types, targets, and memory age shape practice. We reveal clear hierarchy of editability, with identity anchors protected and contexts flexibly modified. We outline applications and risks and propose design implications for responsible GenAI that support remembering.

## Acknowledgments

We extend our deepest gratitude to the participants who voluntarily participated in our study. We also wish to thank Xinyu Chi who helped during the preparation and recruitment stage, and Zehua He for the feedback on earlier versions of this manuscript. We are grateful to all the reviewers for their constructive comments and suggestions, which have improved this work. During the preparation of this work, the authors used ChatGPT 5 and Gemini 2.5 to improve readability and language of the final draft. After using these tools, the authors reviewed and edited the content as needed and take full responsibility for the content of the published article.